\documentclass[%
 aps,
 prb,
 amsmath,amssymb,
 reprint,
]{revtex4-1}

\usepackage{graphicx}% Include figure files
\usepackage{here} %for figure
\usepackage{dcolumn}% Align table columns on decimal point
\usepackage{bm}% bold math
\usepackage{color}

\usepackage{float}  %by AY

\begin{document}

%\preprint{APS/123-QED}

\title{Modulation of probe signal in coherent phonon detection revisited: \\Analytical and first-principles computational analyses}% Force line breaks with \\

\author{Atsushi Yamada}
\email{ayamada@ccs.tsukuba.ac.jp}
% \altaffiliation[Also at ]{Center for Computational Sciences, University of Tsukuba, 1-1-1 Tennodai, Tsukuba, Ibaraki 305-8577, Japan}%Lines break automatically or can be forced with \\
\author{Kazuhiro Yabana}%
\affiliation{%
Center for Computational Sciences, University of Tsukuba, 1-1-1 Tennodai, Tsukuba, Ibaraki 305-8577, Japan 
}%

%\affiliation{
%\affiliation{
%\affiliation{%

\date{\today}% It is always \today, today,

\begin{abstract}
Modulation of probe signal in pump-probe measurements of coherent phonons in dielectrics, 
with and without spectral resolution, are investigated theoretically taking diamond as an example.
Analytical investigation as well as first-principles calculations based on time-dependent 
density functional theory is utilized to clarify the mechanism of the modulation of the probe signals.
Boundary and bulk effects are investigated systematically, 
putting emphasis on the phase relation between the  modulation and the atomic motion of the coherent phonon.
They are summarized as follows:
Modulation by the boundary effect is in phase with the coherent phonon amplitude, while that by
the bulk effect shows $\pi/2$ phase difference. Strong frequency dependence appears in the modulation
by the bulk effect, while no frequency dependence by the boundary effect.
First-principles calculations support the reliability of the analytical result.
\begin{description}
\item[PACS numbers]
xxxxxx
\end{description}
\end{abstract}

\pacs{Valid PACS appear here}% PACS, the Physics and Astronomy
\maketitle

%\tableofcontents

\section{Introduction} \label{sec:intro}

Coherent phonon generation is commonly observed when an intense and ultrashort 
light pulse irradiates on a surface of a bulk material. 
It is usually measured using a pump-probe method in the following way.
A strong pump pulse generates the vibrational motion of atoms in the medium that has a 
coherence in space and time.
A weak probe pulse is then used to detect the coherent phonon through a measurement of
the modulation of the optical response induced by the atomic displacements.
Mechanisms for the generation of coherent phonons in simple crystalline solids have
been extensively discussed since the middle of '80 in both
theoretical and experimental perspectives
\cite{Thomsen1984,Nelson1985,Nelson1987,Cho1990,Merlin1997,Merlin2002,Hase2003,Petek2006,Mizoguchi2013,Nakamura2016,Nakamura2018,
  Nelson1985-2,Nelson1994,Cheng1991,Kato2009,Kitajima2004,Kitajima2010,Sanders2013,
  Pfeifer1992,Scholz1993,Kuznetsov1994,Riffe2007,Glerean2019,Hase2019}.

Recent researches have been extending to further manipulations by,
for example, using multi-pump pulses to control the phonon amplitude\cite{Nelson1994,Nelson2017,Nakamura2018-2},
using stronger pulses to give rise a large amplitude oscillation that may possibly realize
a photoinduced phase transition\cite{Wall2012,Okamoto2017,Horiuchi2017,Schmidt2017,Marieke2017,Fritz2007,Sokolowski2003,Bauerhenne2017}.
Investigations have also been extended to systems other than bulk materials such as
graphene and two-dimensional materials\cite{Kitajima2008,Kitajima2013,Takeda2014}, 
and solids composed of biological molecules\cite{Hase2017}.

In this paper, we devote ourselves to theoretical investigations of the probe stage of pump-probe
measurements of coherent phonons in transparent materials. 
For the generation of coherent phonons in transparent materials, an impulsive stimulated 
Raman scattering (ISRS) mechanism has been widely accepted\cite{Nelson1994,Kitajima2010,Merlin1997}. 
In this mechanism, the pump pulse brings virtual electronic excitations in the medium
during the irradiation that causes the impulsive force acting on atoms.
In the probe stage, modulations of the reflectivity or the transmittivity are usually measured and 
analyzed using a simple and intuitive formula\cite{Merlin1997},
\begin{equation}
\frac{\Delta R}{R} \propto \frac{\partial R}{\partial n} \frac{\partial n}{\partial Q} Q(t),
\label{intuitive}
\end{equation}
where $n$ is the index of refraction and $Q$ is the phonon amplitude.
It can be derived assuming that the modulation takes place at the surface of the medium. 
Using this formula, the modulation is proportional to the phonon amplitude.
We call this mechanism of the modulation the boundary effect below. 
In the probe state, the significance of the bulk effect has also been
discussed immediately after the observation of the coherent phonon
generation by the ISRS mechanism\cite{Nelson1985,Nelson1987}.
It has been pointed out that it shows
a phase shift of $\pi/2$ with respect to the phonon oscillation,
that is, the modulation is maximum when the phonon
amplitude is zero\cite{Nelson1985,Nelson1987,Merlin1997,Liu1995}.
However, the bulk effect has not been observed much
since the effect is suppressed by phase mismatch in practical systems\cite{Merlin1997,Liu1995}.

In measurements of modulations of the probe signal,
spectrally resolved signals have also been reported\cite{Merlin1997,Mizoguchi2013,Nakamura2016,Kitajima2004}.
In these measurements, it has been reported that
the measured spectra show a phase difference of $\pi$
between Stokes and anti-Stokes frequency components that correspond to
above and below the central frequency of the probe pulse, respectively.
In Ref. [\onlinecite{Merlin1997}], it was clearly discussed that the phase difference can be
explained as the bulk effect mentioned above.
In the works afterwards \cite{Mizoguchi2013,Nakamura2016,Kitajima2004}, however, the bulk effect was not discussed.
In Ref. [\onlinecite{Nakamura2016}], instead, it has been argued that the quadratic dispersion
of the Raman tensor is responsible for the modulation.
At present, we consider that it is important to organize the effects that appear
in the probe stage of coherent phonon measurements,
the boundary and the bulk effects and signals with and without spectral resolution.

In this paper, we will investigate the modulation of the probe signal employing two approaches. 
We first discuss an analytic treatment that has been developed previously \cite{Liu1995,Merlin1997}. 
Starting with a propagation equation that describes the probe process of the coherent phonon, 
an approximate analytic solution is constructed. Using the solution, 
analytic formula for the modulation of the reflection and transmission rates are constructed,
separating the boundary and the bulk effects, with and without spectral resolution.
We next present a first-principles computational approach based on time-dependent density functional theory 
(TDDFT)\cite{Runge1984,Ullrich2012}.
We have been developing formalism and computational method to calculate electron dynamics in real time\cite{Yabana1996,Bertsch2000}.
In our previous publication\cite{Shinohara2010,Shinohara2012}, it was shown that the TDDFT is capable 
of describing two generation mechanisms of coherent phonons, ISRS and displacive excitation mechanisms.
Recently, we have extended the theoretical approach so that the propagation of the pulsed light
as well as electronic and atomic motions can be described simultaneously, solving the Maxwell
equation for light propagation, the time-dependent Kohn-Sham equation for electron dynamics,
and the Newton equation in the Ehrenfest dynamics for atomic motions\cite{AYamada2019-2}.
We call it the multiscale Maxwell + TDDFT + MD simulation scheme.
We will use the simulation method to mimic the pump-probe measurement of the coherent phonon
and compare the computational and analytical results.
The simulation method was further extended to combine with polarizable force field model of molecular solids \cite{AYamada2020}

The organization of the present paper is as follows.
In Sec.\ref{sec:analytic}, analytic approach for the modulation of the probe process of coherent phonon 
is developed. In Sec. \ref{sec:simulation}, the first-principles computational approach is explained.
In Sec. \ref{sec:results}, results by the first-principles calculations and by the analytical theory are compared. 
Discussions on previous publications are also given. A summary is presented in Sec.\ref{sec:summary}.

\section{Analitical consideration}   \label{sec:analytic}

\subsection{Setup of the system}  \label{subsec:setup}

We consider a pump-probe measurement of coherent phonon generation in diamond
and focus on the probe stage.
We set the coordinate system such that [100] direction of the cubic diamond crystal structure coincides with the $x$-axis.
The surface of the diamond locates at the $x=0$ plane, a medium in $x>0$ and a vacuum in $x<0$ regions.
We set [010] direction parallel to $y$-axis, and [001] to $z$-axis.

The coherent phonon is assumed to be generated by a pump pulse in the ISRS mechanism
as described below.
The pump pulse is linearly polarized in [011] direction, and propagates along the [100] direction.
The duration of the pulse is much shorter than the period of the optical phonon, and the average 
frequency is much below the bandgap of the diamond.
The pump pulse reaches the surface of the diamond at $t=0$, and propagates with the group
speed of $v_g = c/n_g$ where $n_g$ is the group index of refraction of the diamond.
In the following development, we ignore frequency dependence of the susceptibility.
Therefore, we use the index of refraction $n$ instead of $n_g$ below.
The atomic displacements of the coherent phonon are along [100] direction.

We express the atomic displacement at the position $x$ as
\begin{equation}
\Delta {\bf R}^{\pm} \propto \pm (Q(x,t), 0, 0),
\end{equation}
where the sign $\pm$ indicates that there are two possible directions of the atomic
displacements in the optical phonon.
The phonon displacement $Q(x,t)$ is given by
\begin{equation}
Q(x,t) = \theta(x) q \left( t - \frac{n}{c} x \right),
\end{equation}
where the step function $\theta(x)$ is introduced to indicate the spatial region of the medium.
The function $q(t)$ describes the phonon amplitude at the surface $x=0$.
We assume a sinusoidal form,
\begin{equation}
q(t) = q_0 \sin(\Omega t),
\end{equation}
with the phonon amplitude $q_0$ and the frequency of the optical phonon $\Omega$.
We ignore the damping of the coherent phonon for simplicity.

The coherent phonon induces anisotropy in the refractive index of the diamond
in which the optical axes are given by [011] and [01$\bar 1$] directions.
The anisotropy is measured in time domain using the probe pulse whose duration
is much shorter than the period of the phonon.
As the probe process, we consider the electro-optic(eo) sampling method that has often been used to
detect the signal of the coherent phonon\cite{Cho1990,Pfeifer1992,Hase2003}.
In the method, the probe pulse is linearly polarized along the [010] direction that is
$45^{\circ}$ to the polarization direction of the pump pulse ( [011] direction). 
The Raman scattering wave polarized in [001] direction is then induced
by the interaction between the incident probe pulse and the coherent phonon.
The probe signal is then decomposed into the parallel ([011]) and the perpenducular
([01$\bar 1$]) components with respect to the direction of the pump polarization. 
The difference in the modulations that are recorded in the 
two components provides the information on the coherent phonon. 

We investigate the modulation of the probe signal classifying into four cases:
for reflection and transmission signals with and without spectral resolution.
For the reflected and the transmitted probe pulses, we introduce the frequency-resolved
fluences, $F^{(r)}_{\parallel,\perp}(\omega; \delta)$, and $F^{(t)}_{\parallel, \perp}(\omega; \delta)$,
respectively, where $\parallel, \perp$ indicate parallel and perpendicular components,
$\omega$ is the frequency of the probe pulse and $\delta$ specifies the delay time between the pump and the probe pulses. 
The reflected and transmitted intensities in the absence of the coherent phonon are denoted as
 $F_0^{(r)}(\omega)$ and $F_0^{(t)}(\omega)$, respectively.
The subscript $0$ is used also for other quantities to denote the absence of the coherent phonon.

The spectrally-resolved modulation of the reflectance is defined by
\begin{equation}
  \frac{\Delta R_{\parallel,\perp}(\omega; \delta)}{R_{0}(\omega)}
  = \frac{\Delta F^{(r)}_{\parallel,\perp}(\omega;\delta)}{F^{(r)}_{0}(\omega)},
\end{equation}
where $\Delta R_{\parallel,\perp}$ and $\Delta F_{\parallel,\perp}$ indicate the difference from those without the coherent phonon,
that is, $\Delta R_{\parallel,\perp}=R_{\parallel,\perp} - R_{0}$
and $\Delta F^{(r)}_{\parallel,\perp}=F^{(r)}_{\parallel,\perp} - F^{(r)}_{0}$.
We also introduce a modulation without spectral resolution,
\begin{equation}
  \frac{\Delta R_{\parallel,\perp}(\delta)}{R_{0}}
  = \frac{\int d\omega F^{(r)}_{\parallel,\perp}(\omega;\delta)}{\int d\omega F^{(r)}_{0}(\omega)}.
\end{equation}
The signal of the eo-sampling is then given by 
\begin{equation}
  \Delta R_{eo}(\delta)/R_0 = \left( \Delta R_{\perp}(\delta) - \Delta R_{\parallel}(\delta) \right)/R_0. \label{def-eo}
\end{equation}
We introduce similar quantities for the transmission.

\subsection{Propagation equation}

In order to describe the modulation of the probe pulse, we start from the one-dimensional equation
for light propagation, 
\begin{eqnarray}
  \left( \frac{\partial^2}{\partial x^2} - \frac{1}{c^2}\frac{\partial^2}{\partial t^2} \right) {\boldsymbol E}(x,t)
  = \theta(x) \frac{4\pi}{c^2} \frac{\partial^2 {\boldsymbol P}(x,t)}{\partial t^2}  \label{Maxwell-E}
\end{eqnarray}
where ${\boldsymbol E}(x,t)$ and ${\boldsymbol P}(x,t)$ are the electric field of the probe pulse and the
induced polarization at position $x$ and at time $t$.
The step function $\theta(x)$ indicates that the medium is in $x>0$ region.
For the polarization, we assume a linear and instantaneous relation to the electric field as follows,
\begin{equation}
\left( \begin{array}{c}
P_x \\
P_y \\
P_z
\end{array} \right)
=
\chi
\left( \begin{array}{c}
E_x \\
E_y \\
E_z
\end{array} \right)
+
\frac{\partial \chi_{yz}}{\partial Q}Q
\left( \begin{array}{ccc}
0 & 0 & 0 \\
0 & 0 & 1 \\
0 & 1 & 0
\end{array} \right)
\left( \begin{array}{c}
E_x \\
E_y \\
E_z
\end{array} \right),
\end{equation}
where $\chi$ is the linear isotropic susceptibility at the equilibrium atomic configuration, and
$\partial \chi_{yz}/\partial Q$ is the coefficient of the Raman tensor. We denote
$\partial \chi_{yz}/\partial Q$ as $\chi_R$ below to simplify the formula.
We ignore any retardation effects in the analyses in the following development. It is equivalent to
ignoring the frequency-dependence of $\chi$ and $\chi_R$ in the frequency representation.

Equation (\ref{Maxwell-E}) can be decoupled by introducing parallel and perpendicular 
components of the electric field,
\begin{equation}
E_{\parallel,\perp} = \frac{1}{\sqrt{2}} \left( E_y \pm E_z \right),
\end{equation}
The propagation equations for the $E_{\parallel}$ and $E_{\perp}$ are given by,
\begin{equation}
\frac{\partial^2}{\partial x^2} E_{\parallel,\perp} - \frac{n(x)^2}{c^2} \frac{\partial^2}{\partial t^2} E_{\parallel,\perp}
= \pm \frac{4\pi \chi_R}{c^2} \frac{\partial^2}{\partial t^2} \left[ Q E_{\parallel,\perp} \right],
\label{E_basic_eq}
\end{equation}
where the positive sign $(+)$ for $E_{\parallel}$ and the negative sign $(-)$ for $E_{\perp}$ in the right hand side.
The index of refraction $n(x)$ is given by 
\begin{equation}
n(x) = \left\{ \begin{array}{ll}
1 & (x<0) \\
n & (x>0)
\end{array} \right.
\end{equation}
where $n$ is given by $n=\sqrt{1+4\pi \chi}$. 

In the following, we treat the modulation of the susceptibility caused by the coherent phonon, 
the right hand side of Eq. (\ref{E_basic_eq}), as a perturbation.
First we construct the unperturbed solution ignoring the right hand side of Eq. (\ref{E_basic_eq}).
We express the time profile of the incident electric field as $e^{(i)}(t)$ which is a pulsed field centered at $t=0$.
The unperturbed solution which we denote as $E_0(x,t)$ is given as follows,
\begin{equation}
E_0(x,t) = \left\{ \begin{array}{ll}
e^{(i)} \left( t -\delta - \frac{x}{c} \right) - \frac{n-1}{n+1} e^{(i)} \left( t - \delta + \frac{x}{c} \right), & (x<0) \\
\frac{2}{n+1} e^{(i)} \left( t -\delta - \frac{n}{c}x \right). & (x>0),
\end{array} \right.
\end{equation}
where the center of the incident pulse, $e^{(i)}(t-\delta-x/c)$, is set to arrive at the surface $x=0$ at time $t=\delta$.

We denote the electric field including the perturbed field generated by the coherent phonon as
\begin{equation}
E_{\parallel,\perp}(x,t) = E_0(x,t) + \delta E_{\parallel,\perp}(x,t).
\end{equation}
The perturbed fields, $\delta E_{\parallel,\perp}(x,t)$, satisfy
\begin{equation}
\frac{\partial^2}{\partial x^2} \delta E_{\parallel,\perp} - \frac{n(x)^2}{c^2} \frac{\partial^2}{\partial t^2} \delta E_{\parallel,\perp}
= \pm \frac{4\pi \chi_R}{c^2} \frac{\partial^2}{\partial t^2} \left[ Q E_0 \right].
\end{equation}
As is easily verified, the solution of this equation is given by
\begin{eqnarray}
&&
\delta E_{\parallel,\perp}(x,t) = 
\nonumber\\
&&\left\{ \begin{array}{ll}
\mp \frac{4\pi \chi_R}{n(n+1)^2} q \left( t + \frac{x}{c} \right) e^{(i)} \left( t + \frac{x}{c} - \delta \right)
 & (x<0) \\
\mp \frac{4\pi \chi_R}{n(n+1)^2}  \left\{ 1 + \frac{(n+1)x}{c} \frac{d}{dt} \right\}
q \left( t - \frac{n x}{c} \right) e^{(i)} \left( t - \frac{n x}{c} - \delta \right)
& (x>0)
\end{array} \right.
\label{delta_E}
\end{eqnarray}
We note that the transmitted wave includes the stimulated Raman wave whose amplitude increases
linearly with the propagation distance $x$.

\subsection{Modulation effects}

We evaluate the modulation of the reflectivity in the vacuum region, $x<0$, 
and the modulation of the transmittivity in the medium region, $x>0$.
To evaluate the fluence of the pulse, we utilize the Poynting vector $S(x,t)$
that is given in terms of the electric and the magnetic fields by
\begin{equation}
S(x,t) = \frac{c}{4\pi} E(x,t) H(x,t).
\end{equation}
The fluence of the pulsed light is given as the time integration of the Poynting vector,
\begin{equation}
F(x) = \int dt S(x,t).
\end{equation}
To analyze the frequency component of the fluence, we introduce the spectral decomposition
of the fluence as
\begin{equation}
F(x) = \int_0^{\infty} d\omega F(x,\omega),  \label{Fx}
\end{equation}
\begin{equation}
F(x,\omega) = \frac{c}{4\pi^2} {\rm Re} \left[ E(x,\omega) H^*(x,\omega) \right],
\label{intensity_omega}
\end{equation}
where $E(x,\omega)$ and $H(x,\omega)$ are the Fourier transforms of $E(x,t)$ and $H(x,t)$, respectively.

We first consider the reflectivity and transmittivity in the absence of the coherent phonon
and confirm that we obtain well-known results.
The fluences for the incident, reflected, and transmitted waves that are resolved in frequency,
$F_0^{(i)}(\omega)$, $F_0^{(r)}(\omega)$, $F_0^{(t)}(\omega)$, respectively, are given by
\begin{equation}
F_0^{(i)}(\omega) = \frac{c}{4\pi^2} \vert e^{(i)}(\omega) \vert^2,
\end{equation}
\begin{equation}
F_0^{(r)}(\omega) = \frac{c}{4\pi^2} \left( \frac{n-1}{n+1} \right)^2  \vert e^{(i)}(\omega) \vert^2,
\end{equation}
\begin{equation}
F_0^{(t)}(\omega) = \frac{c}{4\pi^2} \frac{4n}{(n+1)^2} \vert e^{(i)}(\omega) \vert^2,
\end{equation}
where $e^{(i)}(\omega)$ is the Fourier transformation of the incident pulse $e^{(i)}(t)$.
The sign of $F^{(r)}_0$ is defined as positive for the backward direction from the material region.
The frequency-resolved reflectivity and transmittivity are obtained as
\begin{equation}
R_0(\omega) = \frac{F_0^{(r)}(\omega)}{F_0^{(i)}(\omega)} = \left(\frac{n-1}{n+1} \right)^2,
\end{equation}
\begin{equation}
T_0(\omega) = \frac{F_0^{(t)}(\omega)}{F_0^{(i)}(\omega)} = \frac{4n}{(n+1)^2}.
\end{equation}
We note that the reflectivity and transmittivity show no frequency dependence
if we ignore the frequency dependence in the index of refraction.

We next move to the perturbative contributions generated by the coherent phonon.
We calculate them by Eq. (\ref{intensity_omega}) using the electric field of Eq. (\ref{delta_E}) 
and corresponding magnetic field. 
The modulation in the fluence of reflected wave is calculated as
\begin{equation}
\delta F^{(r)}_{\parallel,\perp}(\omega)
= \pm \frac{c \chi_R (n-1) q_0}{\pi n (n+1)^3} {\rm Im} W(\omega,\delta),
\end{equation}
where $W(\omega,\delta)$ is introduced by
\begin{equation}
W(\omega,\delta) = e^{(i)*}(\omega)
\left\{ e^{i\Omega \delta} e^{(i)}(\omega+\Omega) - e^{-i\Omega\delta}e^{(i)}(\omega-\Omega) \right\}.
\end{equation}
From this result, the modulation of the frequency-resolved reflectivity is calculated as
\begin{equation}
\frac{\Delta R_{\parallel,\perp}(\omega,\delta)}{R_{0}(\omega)} = \pm \frac{4\pi \chi_R q_0}{n(n^2-1)} \frac{{\rm Im} W(\omega,\delta)}{\vert e^{(i)}(\omega) \vert^2}.
\label{dR-w}
\end{equation}
The modulation of the reflectivity without the frequency resolution is given by
\begin{equation}
  \frac{\Delta R_{\parallel,\perp}(\delta)}{R_{0}}
  = \pm \frac{8\pi \chi_R}{n(n^2-1)} \frac{\int q(t+\delta)(e^{(i)}(t))^2dt}{\int (e^{(i)}(t))^2dt}. \label{dR-org}
\end{equation}
Assuming that the probe pulse is much shorter than the period of the coherent phonon, $2\pi/\Omega$,
the equation is approximated to simpler form, 
\begin{equation}
\frac{\Delta R_{\parallel,\perp}(\delta)}{R_{0}} \simeq \pm \frac{8\pi \chi_R q_0}{n(n^2-1)} \sin (\Omega\delta).
\label{dR}
\end{equation}
We note that the modulation is in phase with the phonon amplitude and that
this expression coincides with an intuitive expression of Eq.~(\ref{intuitive}) assuming the instantaneous
modulation of the susceptibility given by $\chi(t) = \chi + \chi_R q(t)$ at the surface,
\begin{equation}
  \frac{\Delta R_{\parallel,\perp}(\delta)}{R_0}
  = \frac{1}{R_0}\frac{\partial R}{\partial n_{\parallel,\perp}} \frac{\partial n_{\parallel,\perp}}{\partial Q} q_0 \sin(\Omega \delta).
\end{equation}
The signals of the eo-sampling defined by Eq.(\ref{def-eo}) are then written by
\begin{eqnarray}
  \frac{\Delta R_{eo}(\omega,\delta)}{R_{0}(\omega)}
  &=& - \frac{8\pi \chi_R q_0}{n(n^2-1)} \frac{{\rm Im} W(\omega,\delta)}{\vert e^{(i)}(\omega) \vert^2} \label{dReo(w)R0(w)} \\
  \frac{\Delta R_{eo}(\delta)}{R_{0}}
  &=&    -\frac{16\pi \chi_R}{n(n^2-1)} \frac{\int q(t+\delta)(e^{(i)}(t))^2dt}{\int (e^{(i)}(t))^2dt} \label{dReoR0} \\
  &\sim& -\frac{16\pi \chi_R q_0}{n(n^2-1)} \sin (\Omega\delta).  \label{dReo}
\end{eqnarray}

For the transmittivity, there appear two terms in the modulation of the fluence,
%XXX AY: I exchanged the order of the first and second terms (to be the same order of the later equation)
\begin{eqnarray}
\delta F^{(t)}_{\parallel,\perp}(\omega) 
&=& \mp \frac{c \chi_R}{\pi n (n+1)^3} q_0 {\rm Im} \nonumber\\
&&  \pm \frac{2 \chi_R \omega x}{\pi (n+1)^2} q_0 {\rm Re} W(\omega,\delta) 
    \left[ n W(\omega,\delta) + W^*(\omega,\delta) \right].
\end{eqnarray}
The first term is generated at the surface $x=0$ and 
the second term originates from the stimulated Raman wave which is proportional to the propagation length $x$.
We call the former the boundary contribution and the latter the bulk contribution.
The modulation with the frequency resolution is given by
\begin{eqnarray}
  \frac{\Delta T_{\parallel,\perp}(\omega,\delta)}{T_{0}(\omega)}
  &=& \mp \frac{\pi \chi_R q_0}{n^2(n+1)} \frac{{\rm Im} [nW(\omega,\delta)+W^*(\omega,\delta)]}{\vert e^{(i)}(\omega) \vert^2} \nonumber\\
  &&  \pm \frac{2\pi \chi_R \omega x q_0}{cn} \frac{{\rm Re}W(\omega,\delta)}{\vert e^{(i)}(\omega) \vert^2}.
  \label{dT-w}
\end{eqnarray}
Using Eq.(\ref{Fx}),
the modulation without the frequency resolution is given by
\begin{eqnarray}
  \frac{\Delta T_{\parallel,\perp}(\delta)}{T_{0}}
  &=& \mp \frac{2\pi \chi_R (n-1)}{n^2(n+1)} \frac{\int q(t+\delta)(e^{(i)}(t))^2dt}{\int (e^{(i)}(t))^2dt} \nonumber \\
  & & \mp \frac{2\pi \chi_R x}{cn} \frac{\int \frac{dq}{dt}(t+\delta)(e^{(i)}(t))^2dt}{\int (e^{(i)}(t))^2dt},
\label{dT-org}
\end{eqnarray}
and a simpler expression is obtained by using the short pulse limit approximation as
\begin{equation}
  \frac{\Delta T_{\parallel,\perp}(\delta)}{T_{0}}
  \simeq \mp \frac{2\pi \chi_R (n-1) q_0}{n^2(n+1)} \sin (\Omega\delta)
  \mp \frac{2\pi \chi_R x q_0}{cn} \cos (\Omega \delta).
\label{dT}
\end{equation}
%XXX AY rephrased
From Eqs.(\ref{dT-org}) and (\ref{dT}),
the first sine function terms that originate from the surface come from the phonon amplitude $q(t+\delta)$.
It causes in-phase modulation on the time delay as that in the reflection.
The second cosine function terms that originate from the stimulated Raman wave
are due to the phonon velocity $\frac{dq}{dt}(t+\delta)$
which induces the $\pi/2$ phase shifted modulation.
%%We note that the first term that originates from the surface has
%%the same dependence on the time delay as the modulation in the reflection.
This expression of the first term again coincides with an intuitive expression of Eq.(\ref{intuitive})
assuming the instantaneous modulation of the susceptibility $\chi(t) = \chi + \chi_R q(t)$ at the surface,
\begin{equation}
  \frac{\Delta T^B_{\parallel,\perp}(\delta)}{T_{0}}
  = \frac{1}{T_0}\frac{\partial T}{\partial n_{\parallel,\perp}} \frac{\partial n_{\parallel,\perp}}{\partial Q} q_0 \sin(\Omega \delta),
\end{equation}
where the superscript $B$ indicates that this originates from the boundary effect.
%The second term of Eq.~(\ref{dT}) that originates from the stimulated Raman wave shows the cosine dependence, 
%$\pi/2$ phase shifted from the modulation by the boundary effect.

From Eqs. (\ref{dT-w}), (\ref{dT-org}) and (\ref{dT}),  the transmission change in the eo-sampling are written by
\begin{eqnarray}
  \frac{\Delta T_{eo}(\omega,\delta)}{T_{0}(\omega)}
  &=& - \frac{4\pi \chi_R \omega x q_0}{cn} \frac{{\rm Re}W(\omega,\delta)}{\vert e^{(i)}(\omega) \vert^2}  \nonumber \\
  & & + \frac{2\pi \chi_R q_0}{n^2(n+1)} \frac{{\rm Im} [nW(\omega,\delta)+W^*(\omega,\delta)]}{\vert e^{(i)}(\omega) \vert^2} \\ \label{dTeo(w)T0(w)}
  \frac{\Delta T_{eo}(\delta)}{T_{0}}
  &=& \frac{4\pi \chi_R (n-1)}{n^2(n+1)} \frac{\int q(t+\delta)(e^{(i)}(t))^2dt}{\int (e^{(i)}(t))^2dt} \nonumber \\
  & & + \frac{4\pi \chi_R x}{cn} \frac{\int \frac{dq}{dt}(t+\delta)(e^{(i)}(t))^2dt}{\int (e^{(i)}(t))^2dt} \\   \label{dTeoT0}
  &\simeq&   \frac{4\pi \chi_R (n-1) q_0}{n^2(n+1)} \sin (\Omega\delta) \nonumber \\
  & &      + \frac{4\pi \chi_R x q_0}{cn} \cos (\Omega \delta).
\end{eqnarray}

To simplify the result for the frequency-resolved modulation, we introduce an assumption that 
$e^{(i)}(\omega)$ is a real-valued function except for a multiplicative complex number.
For example, for a symmetric function, $e^{(i)}(t) = e^{(i)}(-t)$, we have real-valued $e^{(i)}(\omega)$.
For an anti-symmetric function, $e^{(i)}(t) = - e^{(i)}(-t)$, $e^{(i)}(\omega)$ is a pure imaginary function.
Under the assumption and expressing $e^{(i)}(\omega)$ removing the complex phase, we have
\begin{eqnarray}
&&W(\omega,\delta) =
\cos(\Omega\delta) e^{(i)}(\omega) \left\{ e^{(i)}(\omega+\Omega) - e^{(i)}(\omega-\Omega) \right\}
\nonumber\\
&& \hspace{3mm} + i \sin(\Omega\delta) e^{(i)}(\omega) \left\{ e^{(i)}(\omega+\Omega) + e^{(i)}(\omega-\Omega) \right\}.
\end{eqnarray}
We may further introduce an expansion with respect to $\Omega$ that is justified when the probe
pulse is much shorter than the period of the coherent phonon. Then $W(\omega,\delta)$ is approximated as
\begin{equation}
W(\omega,\delta) \simeq \cos(\Omega\delta) \Omega \frac{d}{d\omega} \left( e^{(i)}(\omega) \right)^2
+ i \sin(\Omega\delta) \left( e^{(i)}(\omega) \right)^2.
\end{equation}
Using this approximation, we get the following simplified expressions for the spectrally-resolved modulations,
%XXX AY fixed the coefficients
\begin{equation}
\frac{\Delta R_{eo}(\omega,\delta)}{R_{0}(\omega)}
\simeq - \frac{8\pi \chi_R q_0}{n(n^2-1)} \sin \Omega\delta,
\label{dR-w-a}
\end{equation}
\begin{eqnarray}
\frac{\Delta T_{eo}(\omega,\delta)}{T_{0}(\omega)}
&\simeq&  - \frac{4\pi \chi_R \omega \Omega x q_0}{cn} \cos \Omega\delta
\frac{\frac{d}{d\omega} \left( e^{(i)}(\omega) \right)^2}{\left( e^{(i)}(\omega) \right)^2}
\nonumber\\
&&
+ \frac{2 \pi \chi_R (n-1) q_0}{n^2(n+1)} \sin \Omega \delta.
\label{dT-w-a}
\end{eqnarray}
We note that the terms originated from the boundary, the modulation of the reflection and
the second term of the modulation of the transmission, are independent of the frequency
in the first order approximation,
while the term originated from the stimulated Raman wave causes the frequency-dependent
modulation.

We here mention relation of our results with previous works.
The bulk effect as well as the boundary effect was discussed by Merlin and
collaborators \cite{Merlin1997, Liu1995}.
%XXX AY (The same or similar)
Corresponding expressions to Eqs. (\ref{dR-w}) and (\ref{dT-w}) were presented there.
In particular, the appearance of the frequency dependence as well as the phase
change between the bulk and the boundary effects have been stressed.
In the present derivation, we provide a unified and detailed explanation of the formula 
with a precise expression for the amplitude of the  modulation, 
which was not presented in Ref. \onlinecite{Merlin1997}.
In Ref. [\onlinecite{Nakamura2016}],
it was argued that the frequency-dependent modulation cannot be described without introducing a frequency-dependence in the Raman tensor.
This conclusion contradicts with the present result: the frequency-dependent modulation can be explained if we include the bulk effect.

\section{First-principles simulation based on time-dependent density functional theory} \label{sec:simulation}

\begin{figure}[t]
\centering
\includegraphics[angle=0,width=0.98\linewidth]{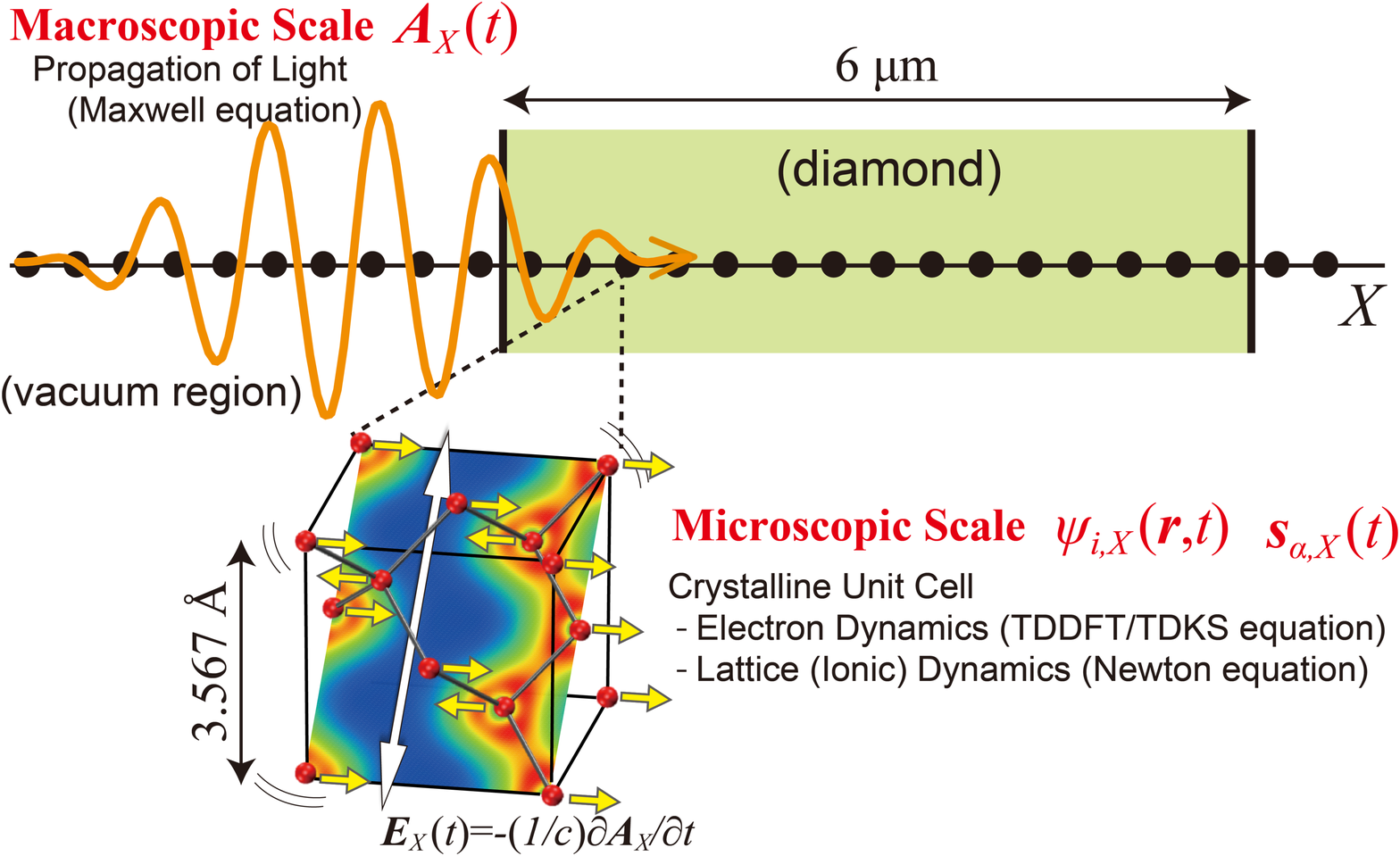}
\vspace{-4mm}
\caption{Schematic illustration of the multiscale model.
}
\label{fig-multiscale}
\end{figure}

In the previous section, we used several assumptions and approximations to derive the analytical formula.
For example, we ignored the frequency dependence of the Raman tensor as well as the dielectric function.
We also take a lowest order perturbation theory.
Harmonic motion is assumed for the phonon motion.
In this section, we present a complimentary computational approach based on first-principles 
time-dependent density functional theory.
We develop a multiscale formalism\cite{Yabana2012} that allows description of the pump-probe measurement of coherent
phonons without any empirical parameters related to materials\cite{AYamada2019-2}. 
%XXX AY.  In this section, we present a formalism and numerical method.
A formalism and numerical method are written in this section and 
calculated results will be presented and compared with analytical results in the next section.

\subsection{Multiscale Simulation method}
Here we briefly explain our multiscale simulation method.
A full explanation has been given in our previous publication\cite{AYamada2019-2}.
Calculations have been carried out using SALMON, an open source software developed in our group\cite{SALMON_paper2018,SALMON_web}.

Our simulation describes the pump-probe measurement of the coherent phonon generation faithfully 
mimicing the setup of the problem.
We show the scheme of our simulation in Fig. \ref{fig-multiscale}.
For an incident pulse propagating along the $x$ axis, 
we utilize two coordinate systems: The light propagation is described using a one-dimensional coordinate $X$,
which we call the macroscopic coordinate.
Microscopic three-dimensional coordinates $\bm r$ is used to describe the dynamics of electrons and ions.

The light electromagnetic field is expressed by using a vector potential
${\bm A}_X(t)$. It satisfies the Maxwell equation in the macroscpic scale,
\begin{equation}
\left[ \frac{1}{c^2} \frac{\partial^2}{\partial t^2} - \frac{\partial^2}{\partial X^2} \right] {\bm A}_X(t)
= \frac{4\pi}{c} {\bm J}_X(t),  \label{Maxwell}
\end{equation}
where ${\bm J}_X(t)$ is the electric current density at the point $X$.

In solving Eq.~(\ref{Maxwell}), we discretize the coordinate $X$ using a uniform grid.
At each macroscopic grid point $X$, we consider a microscopic dynamics of electrons and ions.
In our multiscale description, each microscopic dynamics is assumed to be regarded as infinitely periodic.
Since the wavelength of the pulsed light is much longer than the typical spatial scale of
the microscopic dynamics, we assume a dipole approximation where electrons and ions move 
under a spatially-uniform electric field, ${\bm E}_X(t) = -(1/c)(\partial {\bm A}_X(t)/\partial t)$.
Then we may apply the Bloch theorem in the microscopic dynamics: The electron motion at macroscopic position $X$ is 
described using Bloch orbitals $u_{n {\bm k},X}({\bm r},t)$ specified by the macroscopic position $X$, 
band index $n$, and the crystalline momentum ${\bm k}$.
Ionic motion is described by the coordinates of ions in the unit cell, ${\bm R}_{\alpha,X}(t)$, 
where the index $\alpha$ distinguishes different ions in the unit cell.
 
The Bloch orbitals satisfy the TDKS equation,
\begin{eqnarray}
&& i\hbar \frac{\partial}{\partial t} u_{n{\bm k},X}({\bm r},t)= \nonumber\\
&& \left[ \frac{1}{2m} \left\{ -i\hbar \bm \nabla_{\bm r} + \hbar{\bm k} + \frac{e}{c} {\bm A}_X(t) \right\}^2
-e \phi_X({\bm r},t) \right.  \nonumber\\
&& \left. + \frac{\delta E_{XC}[n_{e,X}]}{\delta n_{e,X}}
+ \hat v_{{\rm ion},X}(\bm r,t) \right] u_{n{\bm k},X}({\bm r},t) \label{TDKS} 
\end{eqnarray}
where 
$n_{e,X}$ is the electron density given by $n_{e,X}({\bm r},t)=\sum_{n,{\bm k}}|u_{n{\bm k},X}({\bm r},t)|^2$.
$\phi_X({\bm r},t)$ and $E_{XC}[n_{e,X}]$ are the Hartree potential and the exchange-correlation energy, respectively. 
$\hat v_{{\rm ion},X}(\bm r,t)$ is the electron-ion potential for which we use
norm-conserving pseudopotential \cite{Troullier1991}. 
%$\phi_X$ and 
The ionic potential $\hat v_{{\rm ion},X}$ depends on the ionic coordinates $\{\bm R_{\alpha,X}(t)\}$ as parameters.

To describe the dynamics of ions, we use a so-called Ehrenfest method \cite{Ullrich2012}
where the ionic motion is described by the Newtonian equation,
\begin{eqnarray}
  M_{\alpha} \frac{d^2 \bm R_{\alpha,X}}{dt^2}
  =-\frac{eZ_{\alpha}}{c} \frac{d\bm A_X}{dt}
  -\frac{\partial}{\partial \bm R_{\alpha,X}}
\int d\bm r [en_{{\rm ion},X}\phi_X]  \label{Newton} 
\end{eqnarray}
where $M_{\alpha}$ is the mass of the $\alpha$-th ion, $n_{{\rm ion},X}$ is the charge density of ions given by
$n_{{\rm ion},X}({\bm r},t)=\sum_{\alpha}Z_{\alpha}\delta({\bm r}-{\bm R_{\alpha,X}}(t))$, 
with $Z_{\alpha}$ the charge number of the $\alpha$-th ion. 

The electric current density at point $X$, $\bm J_X(t)$, consists of electronic and ionic contributions,
\begin{equation}
\bm J_X(t) = \bm J_{e,X}(t) + \bm J_{{\rm ion},X}(t).
\label{current}
\end{equation}
The electronic component $\bm J_{e,X}(t)$ is expressed in terms of the Bloch orbitals $u_{n\bm k,X}(\bm r,t)$ \cite{Yabana2012},
and the ionic component $\bm J_{ion,X}(t)$ is given by the velocity of the ion, $(d/dt){\bm R}_{\alpha,X}(t)$.
  
We solve Eqs. (\ref{Maxwell}) - (\ref{current}) simultaneously to obtain the whole dynamics at once.
The initial condition is so prepared that the electronic state at each macroscopic point $X$ is set to the ground state solution of
the static density functional theory, the ionic positions are set to their equilibrium positions in the electronic 
ground state, and the vector potential of the incident pump- and probe-pulsed light is prepared in the vacuum region 
in front of the film.

We note that the light propagation equation Eq. (\ref{Maxwell-E}) can be identified with Eq. (\ref{Maxwell}),
if we make several assumptions and approximations.
if we make several assumptions and approximations.
They includes:  %For the ionic motion,
We need to assume that the amplitude of the ionic motion is sufficiently small.
The amplitude of the incident pulsed light needs to be sufficiently small so that any nonlinear optical effects other than the
Raman process can be ignorable. We also need to assume that there is no retardation effects in the electronic response
that are equivalent to ignoring the frequency dependence of the susceptibilities.
We will compare the first-principles calculations and the analytic formula to assess the validity of the approximations 
that are required to derive the analytic formula in the previous section.

%--------------------------------------------

\subsection{Computational Details}

We carry out the simulation in the setting of the eo-sampling.
As the time profiles of the incident pump and probe pulses, we choose cosine-squared shaped 
envelope given as
%XXX AY: revised the time region
\begin{eqnarray}
  \bm A_{\rm pump}(t)  &=& A_{\rm pump} \cos^2 \left(\frac{\pi t}{T} \right) \cos( \omega_0 t) {\bm e_{011}} \\
  \bm A_{\rm probe}(t) &=& A_{\rm probe}\cos^2 \left(\frac{\pi t}{T} \right) \cos( \omega_0 t) {\bm e_{010}} \\
  & & { } \hspace{10mm} (-T/2 < t < T/2)  \nonumber
\end{eqnarray}
where $\bm e_{011}(=\bm e_{yz}=\left( \bm e_y + \bm e_z \right)/\sqrt{2})$ and 
$\bm e_{010}(=\bm e_{y})$ are the spatial unit vectors of the polarization direction of the pump and probe pulses, respectively. 
The incident probe pulse is given as $\bm A_{\rm probe}(t-\delta)$ with the pump-probe delay time $\delta$.
The delay time is chosen to be 83.0, 89.5 and 96.0 fs.
The average frequency $\omega_0$ of the pump and the probe pulses are chosen to be a common value, $\hbar\omega_0$=1.55eV.
The pulse duration of $T$=18 fs is used for all pulses. This amounts to the pulse duration of 7 fs in FWHM. 
It is much shorter than the period of the optical phonon of diamond that is about 25 fs.
The incident intensities of the pump and the probe pulses are set to $2\times 10^{12}$ W/cm$^2$ and $1\times 10^{10}$ W/cm$^2$, respectively.
At these intensities, nonlinear electronic excitations across the bandgap is not significant.

In practical calculations, we carry out calculations of the pump and the probe processes separately.
In the pump stage, we calculate the propagation of the pump pulse in the medium of thickness 10 $\mu$m and the duration of 80 fs.
At the final time, the pump pulse stays in the spatial region 6 $\mu$m $< X <$ 10 $\mu$m.
In the probe stage, we prepare a diamond medium in the spatial region of 0 $\mu$m $< X <$ 6 $\mu$m.
In this spatial region, the initial ionic motions is prepared from the coherent phonon obtained in the pump stage calculation.

%XXX AY: calculation conditions? (not only parameters) -> calculation system and parameters ?
The calculation system and parameters are the same as those of Ref.\onlinecite{AYamada2019-2}.
The macroscopic coordinate $X$ is discretized using the spacing of 15nm.
In the microscopic calculation, adiabatic local density approximation\cite{Perdew1981} is used for the exchange-correlation potential.
The unit cell consisting of eight carbon atoms in the cubic cell with the side length of 3.567 {\AA} is used.
The Bloch orbitals are expressed using 16$^3$ uniform spatial grids in the unit cell and 12$^3$ of k-points in the Brillouin zone.
All of the equations of motion are integrated with a common time step of 0.02 fs.

\section{Calculated results}   \label{sec:results}
\vspace{-3mm}

The purpose of this section is to compare the results between the analytical description developed in
Sec. \ref{sec:analytic} and the first-principles simulation described in Sec. \ref{sec:simulation}.
Although analytical treatments provide formula that are useful to understand mechanisms of probe
process, several assumptions and approximations are used in the derivation including the perturbative
expansion and the ignorance of the frequency-dependence of the response. 
Contrarily, the first-principles calculation does not require these assumption and approximation.
Therefore, the comparison between two approaches will be useful to assess the validity of the 
analytical approach, and to clarify the significance of the effects that are not included in the analytical approach.
%In particular, it has been reported that the frequency-dependence of the susceptibility may explain
%the frequency-dependent modulation in the probe signal of coherent phonon\cite{Nakamura2016}.
%To clarify the origin of the frequency-dependent modulation, it will be important to compare the first-principles
%calculation that includes both effects of stimulated Raman wave and frequency-dependent susceptibilities.

When we make numerical evaluation of Eqs. (\ref{dR-w}), (\ref{dR}), (\ref{dT-w}), and (\ref{dT}),
we use the following values for the parameters that are chosen to fit the first-principles TDDFT calculation:
$n$=2.25, $\Omega=2\pi/{\rm (25.47 [fs])}$, $q_0$=0.95$\times$10$^{-4}$[\AA], and $\chi_R$=11.6 [\AA$^{-1}$].
The values of $q_0$ and $\chi_R$ are so determined that the generation of the coherent phonon is described
consistently by solving Eq. (\ref{Maxwell-E}) numerically in the presence of the pump pulse.

\subsection{Generation of coherent phonon by pump pulse}
\vspace{-3mm}

\begin{figure}[t]
\centering
\includegraphics[angle=0,width=0.98\linewidth]{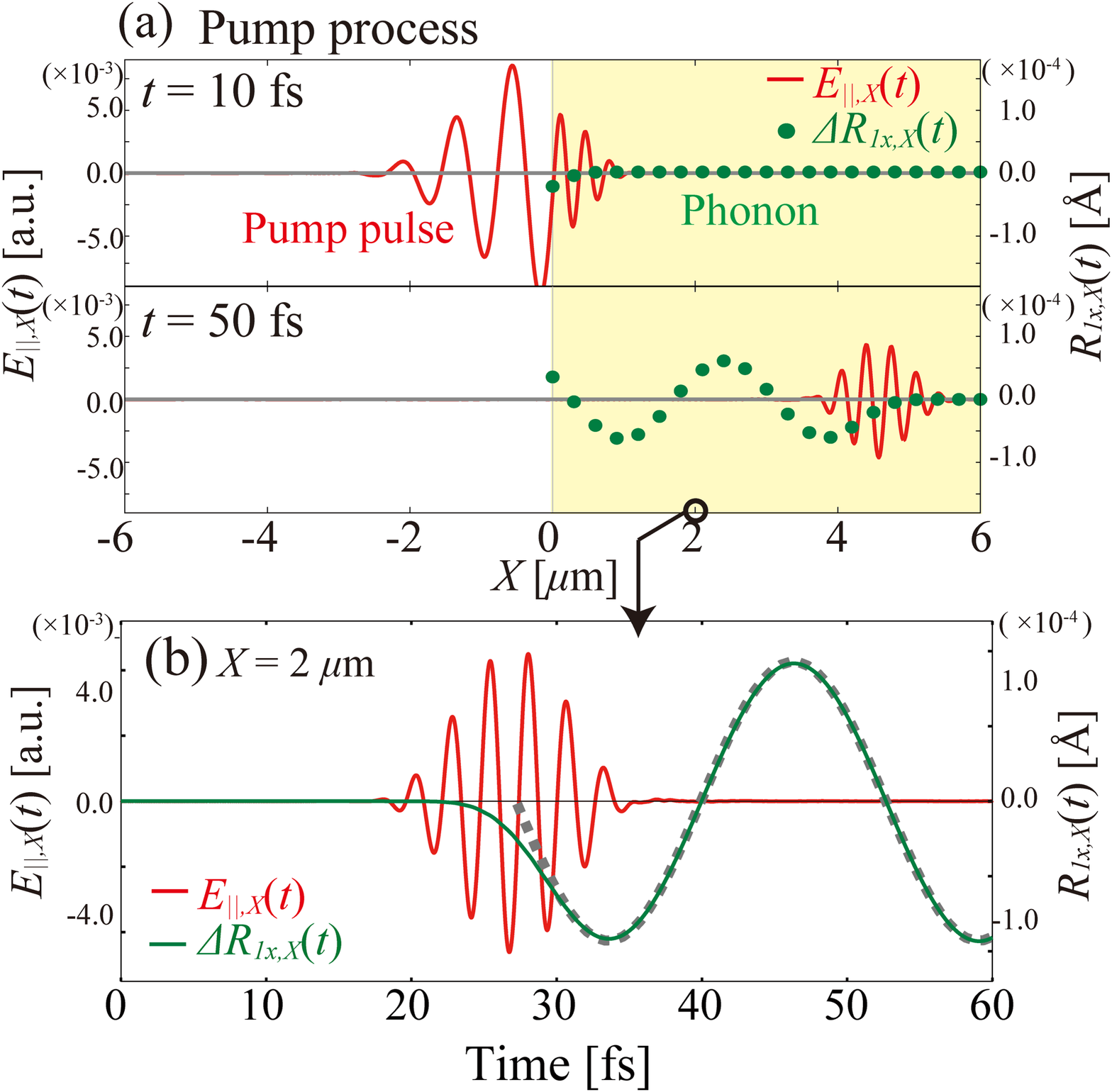}
\vspace{-4mm}
\caption{
Electric field and atomic displacement in the pump stage.
}
\label{fig-pump}
\end{figure}

In the analytical approach, we simply assumed a sinusoidal time profile of the coherent phonon that
propagates with the group velocity of the pump pulse in the medium. We first compare the
time profile of the coherent phonon calculated by our multiscale simulation with the assumed one.
Figure \ref{fig-pump} shows a comparison. In panel (a), the electric field and the atomic displacement
are shown at two times, $t=10$ fs when the pulse just arrived the surface and $t=50$ fs when the pulse
propagates at about $X=4.5$ $\mu$m. In panel (b), the atomic displacement at $X=2$ $\mu$m is
shown as a function of time. The calculated displacement shown by solid curve is well fitted by a
sinusoidal function that is shown by dashed curve. To compare the phonon period and the pulse
duration, the time profile of the incident pulse is shown by red solid curve.
The first-principles calculation includes various nonlinear, non-perturbative, and frequency-dependent
effects. For example, the pump pulse may excite electrons by multiphoton excitation processes
that may cause decrease of the coherent phonon amplitude as the pump pulse propagates.
Anharmonicity in the atomic motion may also affect the time-dependence of the phonon amplitude.
However, the comparison indicates that the simple ISRS mechanism describes accurately the
production stage of the coherent phonon in the present setting of the multiscale calculation.

\subsection{Modulation in transmission}  \label{subsec:Tchange}
\vspace{-3mm}

We move to the probe process.
We first consider the modulation on the transmission.
In the first-principles calculation, we analyze the transmitted wave that appears 
in the vacuum region right to the back surface. There may appear delayed transmitted waves 
that experience internal reflections inside the medium. Since we stop our calculation
when the end of the first transmitted wave passes through the back surface, however,
we do not take into account these waves of multiple reflections at the surfaces.

\begin{figure}[tb]
\centering
\includegraphics[angle=0,width=0.98\linewidth]{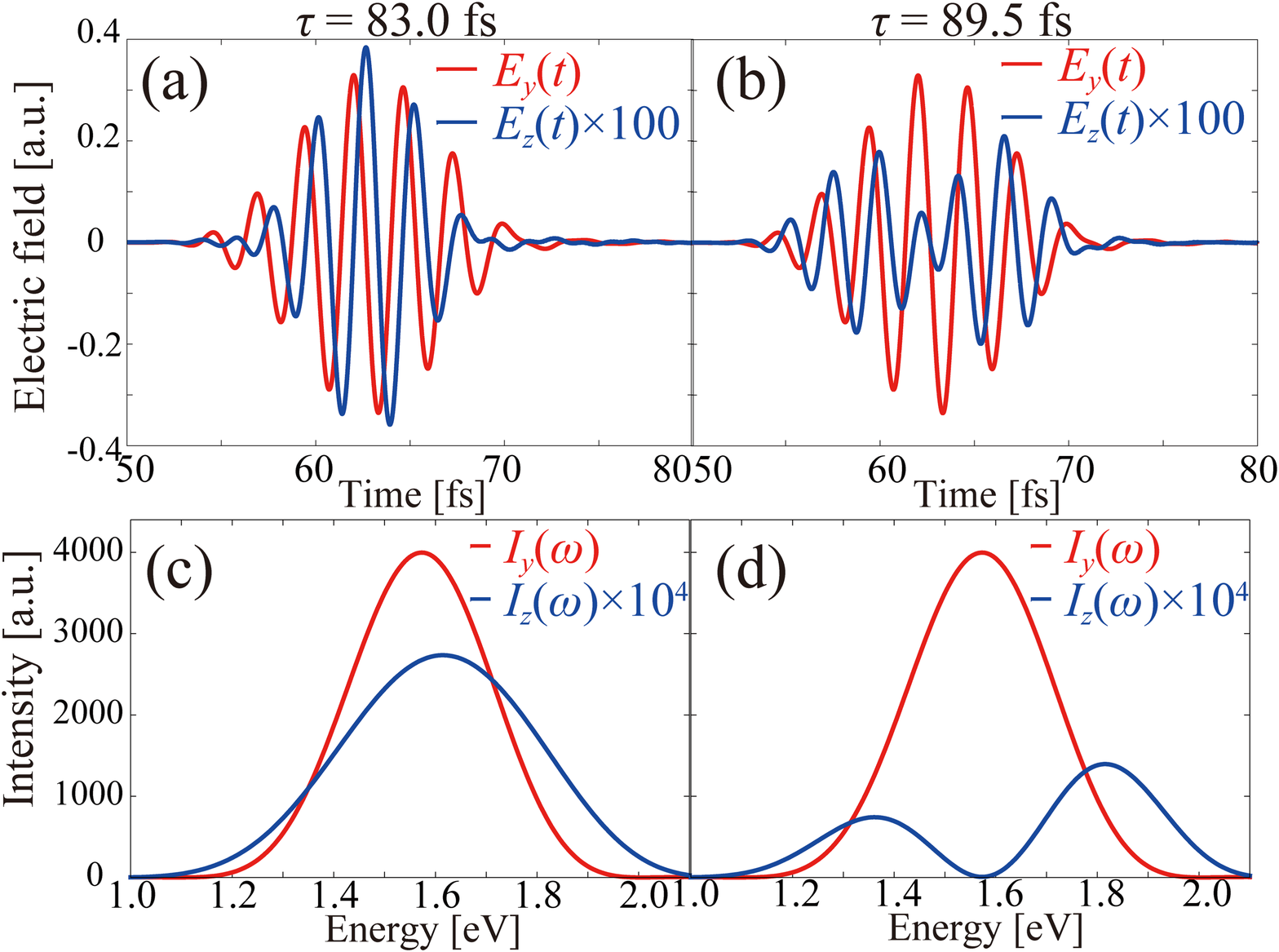}
\vspace{-4mm}
\caption{
  The transmitted electric field in $y$- and $z$-directions calculated by the first-principles simulation, 
  corresponding to the transmitted probe wave and stimulated Raman wave, respectively, 
  obtained in the right vacuum region for (a) $\delta$=83.0 fs and (b) 89.5 fs, 
  and their Fourier transformed power spectra for (c) $\delta$= 83.0 fs and (d) 89.5 fs.
}
\label{fig-Et}
\end{figure}

We first show shapes of the transmitted waves.
Fig. \ref{fig-Et}(a) and (b) show the transmitted electric field for two different pump-probe time delay, 
$\delta$=83.0 fs and 89.5 fs, respectively.
In the former case of $\delta$=83.0, the probe pulse arrives at the surface of the diamond 
when the phonon amplitude is the maximum.
In the latter case of  $\delta$=89.5 fs, the probe pulse arrives at the surface when the phonon amplitude
shows the node.

In our first principles calculation, we employ the probe pulse with the polarization in $y$-direction.
During the propagation, the stimulated Raman wave grows linearly with the propagation length
and appears as the $z$-component of the field. 
In the notation of Sec. \ref{sec:analytic}, the $z$-component of the electric field $E_z(x,t)$ is
equal to $\sqrt{2} \delta E_{\parallel}(x,t)$.
As is shown in Eq.~(\ref{delta_E}), the transmitted wave is composed of two terms, the boundary
term that is created at the surface of the medium and the bulk term that is linearly proportional
to the propagation distance.
The relative significance of the two terms depends on the duration of the probe pulse, phonon frequency, 
and propagation distance. In the present setting with the propagation distance of 6 $\mu$m,
the bulk contribution is much more dominant than the boundary contribution.
Therefore, we expect the form,
\begin{equation}
E_z(x,t) \propto x \frac{d}{dt}  \left[ q \left(t-\frac{nx}{c} \right) e^{(i)} \left( t-\frac{nx}{c} -\delta \right) \right].
\end{equation}

\begin{figure}[htb]
\centering
\includegraphics[angle=0,width=0.98\linewidth]{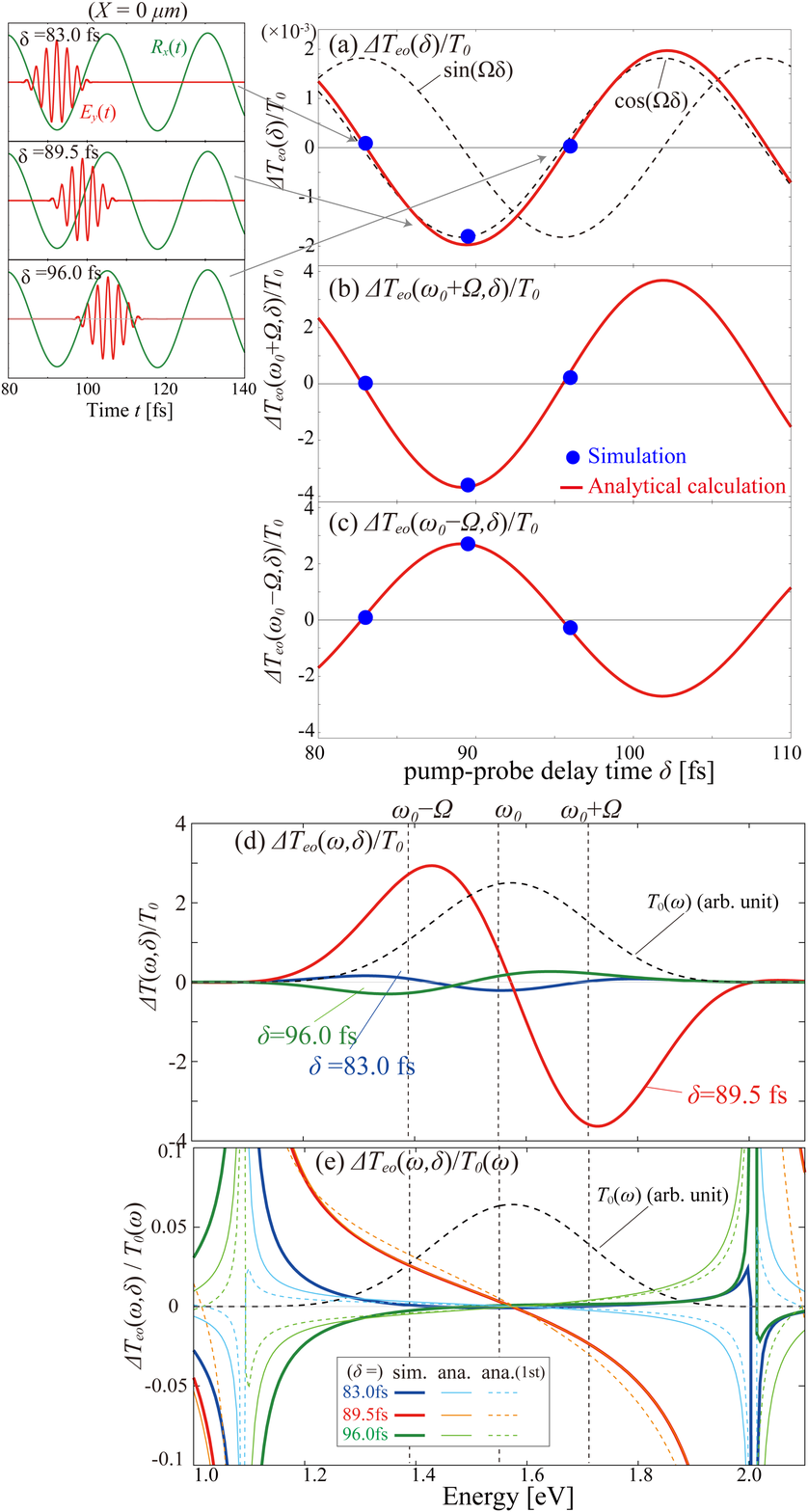}
\vspace{-4mm}
\caption{
  (a) Transmission change ($\Delta T_{eo}(\delta)/T_0$) and
  the frequency components at (b) anti-Stokes and (c) Stokes positions ($\Delta T_{eo}(\omega_0 \pm \Omega,\delta)/T_0$, respectively)
  as a function of delay time $\delta$
  obtained by the simulation(the blue filled circles) and the analytical calculations of
  Eqs. (\ref{dTeoT0}) and (\ref{dTeo(w)T0(w)}) (but $T_0$ is used insted of $T_0(\omega)$) (the red lines).
  The insets in (a) show the probe pulses (the red lines) and atomic displacement (the green lines) as a function of $t$ at $X$=0 $\mu$m in the simulation for each $\delta$.
  Spectrally resolved transmission change as a function of $\omega$: (d) $\Delta T_{eo}(\omega,\delta)/T_0$, obtained by the simulation 
  and (e) $\Delta T_{eo}(\omega,\delta)/T_0(\omega)$ by the simulations and analytical calculations (Eqs.(\ref{dTeo(w)T0(w)}) and (\ref{dT-w-a})). 
}
\label{fig-dT}
\end{figure}

The pulse shapes in Fig. \ref{fig-Et}(a) and (b) indeed show the expected behavior.
In the case of $\delta$=83.0 fs, the envelope shape of the stimulated Raman wave is similar to that of the transmitted probe pulse.
There is a phase shift of $\pi/2$ between the incident ($E_y(t)$) and the stimulated Raman ($E_z(t)$) waves.
This is understood as follows: Since the duration of the probe pulse is shorter than the period of the phonon,
the product $q(t)e^{(i)}(t)$ is mostly proportional to $e^{(i)}(t)$. The time derivative causes the phase shift of $\pi/2$.
In the case of $\delta$=89.5 fs the shape of the stimulated Raman wave is very different from the probe pulse.
This is because the probe pulse propagates with the nodal point of the phonon so that the product $q(t) e^{(i)}(t)$
behaves approximately as $q(t) e^{(i)}(t) \propto t e^{(i)}(t)$. This explains the nodal behavior in the Raman wave
at $\delta$=89.5 fs. 

In panels (c) and (d) of Fig.\ref{fig-Et}, frequency-resolved intensities are shown for $E_y(t)$ and $E_z(t)$.
At $\delta$=83.0 fs, the spectrum of the Raman wave shows somewhat a wider distribution and is slightly
shifted to the higher frequency. At $\delta$=89.5 fs, the spectrum of the Raman wave shows a double-peak
structure. This originates from the extra node in the time domain.

In Fig.\ref{fig-dT}(a), transmission changes without frequency resolution in the first-principles calculation are shown by dots for three 
different pump-probe delay time, $\delta$=83.0, 89.5, and 96.0 fs. The delay time $\delta$=96.0 fs corresponds
to the arrival of the probe pulse at the maximum of the phonon amplitude, as in $\delta=83.0$ fs.
The modulation is large at $\delta$=89.5 fs and very small at $\delta$ = 83.0 fs and 96.0 fs.
%XXX AY changed: please check
This is reasonable according to Eq.(\ref{dT-org}) since the phonon velocity is maximum at $\delta$ = 89.5 fs at the nodal point.
%This is reasonable since the change of the pulse shape is maximum at $\delta$ = 89.5 fs as is seen in Fig.\ref{fig-Et}(b).
The modulation of the transmission signal calculated by the first-principles calculation coincides accurately 
with the analytics formula of Eq.~(\ref{dT}) that shows a cosine-like dependence.
%XXX AY moved here
For the plots of the analytical calculations, 
since we consider the transmitted wave that appears in the right vacuum region, we multiply
a factor of 2 for the term of the boundary effect to take into account the transmission through two boundaries.

We thus find a satisfactory coincidence between results of the analytic theory and those by the first-principles calculation.
This fact indicates that the analysis based on the model presented in Sec.\ref{sec:analytic}
is sufficient to describe the modulation of the transmission signal.
Namely, the modulation in the present case that is dominated by the stimulated Raman wave
can be accurately described using the classical model of the light propagation with harmonic oscillator approximation for the phonon motion,
first-order expansion in the light intensity,
and ignorance of dispersion effects in both the diagonal dielectric function and the off-diagonal Raman tensor.
The modulation is caused mainly by the stimulated Raman wave in the present case.

In panels (b) and (c) of Fig.\ref{fig-dT},
modulations at the anti-Stokes and Stokes frequencies, $\omega_0 \pm \Omega$
are shown as a function of the pump-probe delay time, $\delta$.
The first-principles and the analytical calculations show again excellent agreement.
%The bulk contribution terms in the equations show that
%the double-peak structure comes from the derivative of the power spectrum of the incident probe pulse.
It is noted that the signal shows a striking phase difference between two components at anti-Stokes and Stokes frequencies.
%The phase difference of $\pi$ between the plots of the Stokes and anti-Stokes frequency components
%corresponding to the large positive and negative peak in Fig. \ref{fig-dT}(a)
%are coming from the derivative of the power spectrum.
%the double-peak structure.

%If the signal has a sine curve with the phase difference of zero to the phonon,
%it can be easy to intuitively understand that the signal is larger when the incident pulse interacts with phonon at larger atomic displacement. 
%Our results of the cosine curve signal, however, show that
%the signal has near zero when the incident pulse interacts with the phonon at maximum atomic displacement,
%and the signal becomes maximum when the pulse interacts at the nodal position of the phonon.

To investigate the frequency-dependent modulation in detail, frequency-resolved modulation of the
transmission is shown in Fig. \ref{fig-dT}(d) and (e) for three cases of delayed time, $\delta$=83.0, 89.5, and 96.0 fs
at which the signals are plotted in (a) - (c). 
As seen in panel (d), the modulation is maximum at $\delta=89.5$ fs when the probe pulse moves
with the nodal point of the coherent phonon. It also shows a phase change across approximately
the central frequency of the probe pulse, 1.55 eV.
The modulation is rather small at the delay times of $\delta$=83.0 and 96.0 fs.
%XXX AY changed: please check
These findings are consistent with the first term of Eq. (\ref{dT-w-a}) that shows the differential of the spectrum of the incident pulse.
% These findings are consistent with the pulse shape of the Raman wave shown in Fig.\ref{fig-Et}.

In panel (e), the modulation divided by the frequency-resolved transmission is shown.
This is the quantity often analyzed in experimental analyses.
Here three lines are shown for each pump-probe delay time. 
Solid thick lines show the first-principles calculation, solid thin lines show analytical results
using Eq. (\ref{dT-w}), and thin dashed lines show approximate analytical results using Eq. (\ref{dT-w-a}).
%XXX AY:(this sentence is moved to other paragraph)
%Since we consider the transmitted wave that appears in the right vacuum region, we multiply
%a factor of 2 for the term of the boundary effect to take into account the transmission through
%two boundaries.
The signal is again strong at $\delta=89.5$ fs, The modulation shows a nodal structure around the
average frequency of the pulse and becomes larger as the frequency apart from the average frequency.
The first-principles calculation coincides accurately with the analytic formula of Eq. (\ref{dT-w}).
The simplified analytic formula of Eq. (\ref{dT-w-a}) somewhat deviates from others. The difference
is not very significant.
At $\delta$=83.0 and 96.0 fs, the signal is small for all frequencies. As the frequency comes apart from
the central frequency of the probe pulse, the signal becomes larger. However, the signal
showing divergent behavior at frequencies around 1.1 eV and 2.0 eV will not be physically significant
since the component of the probe pulse in those frequency region is extremely small.
Looking at the panel (e) in detail, the signal at $\delta$=96.0 fs shows a negative (positive) modulation
at low (high) frequency region in both first-principles and analytic results. However, at $\delta$=83.0 fs,
though the analytic formula suggest opposite behavior while the first-principles calculation shows
positive modulation in both side. We do not have an explanation for this observation.

\subsection{Modulation in reflection}  \label{subsec:Rchange}
\vspace{-3mm}

\begin{figure}[ht]
\centering
\includegraphics[angle=0,width=0.98\linewidth]{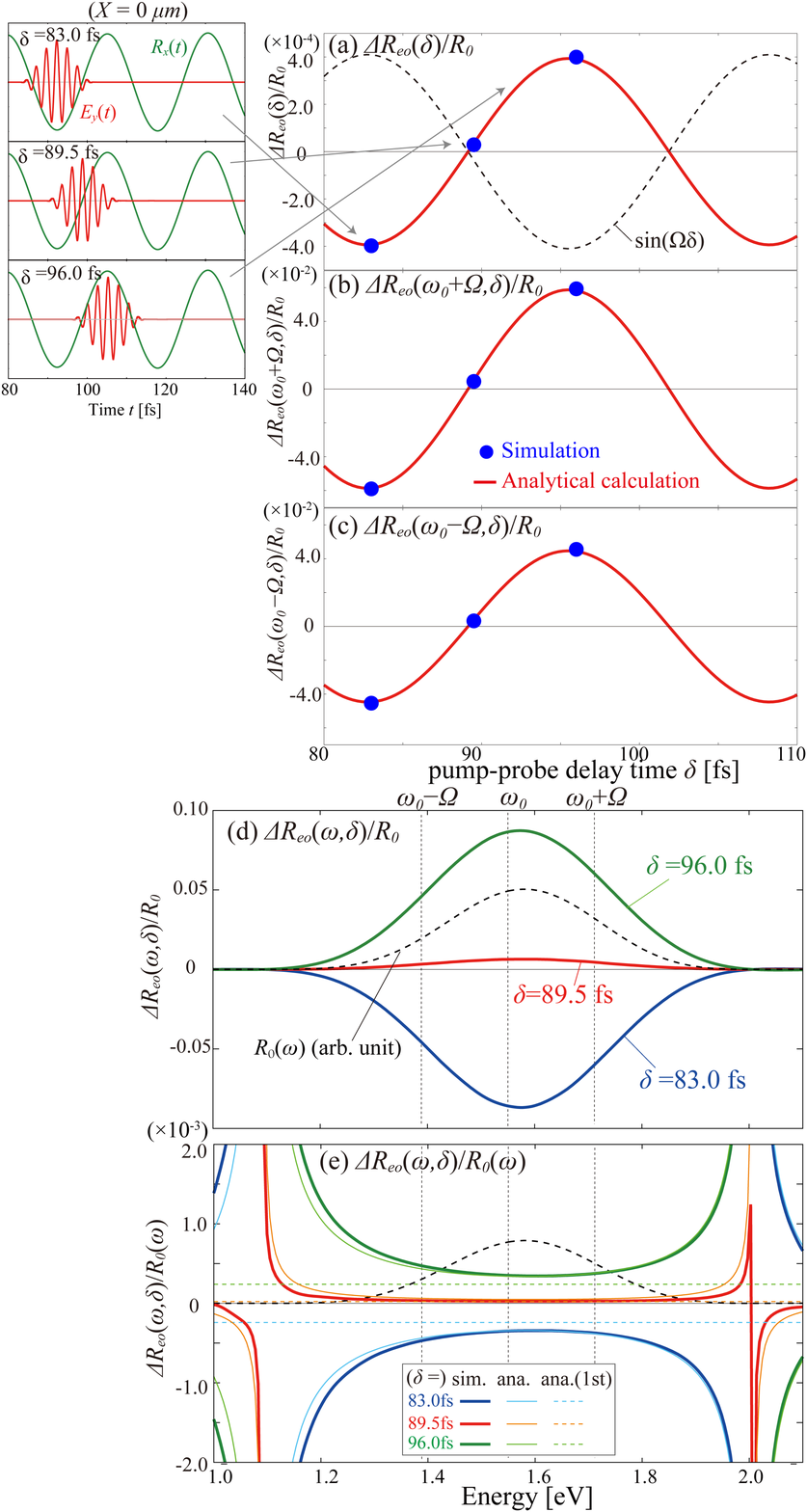}
\vspace{-4mm}
\caption{
  (1) Reflection change ($\Delta R_{eo}(\delta)/R_0$) and
  the frequency components at (b) anti-Stokes and (c) Stokes positions ($\Delta R_{eo}(\omega_0 \pm \Omega,\delta)/R_0$, respectively)
  as a function of delay time $\delta$
  obtained by the simulation(the filled blue circles) and the analytical calculations of
  Eqs. (\ref{dReo(w)R0(w)}) and (\ref{dReoR0}) (but $R_0$ is used instead of $R_0(\omega)$) (the red lines).
  The insets in (a) show the probe pulses (the red lines) and atomic displacement (the green lines) as a function of $t$ at $X$=0 $\mu$m in the simulation for each $\delta$.
  Spectrally resolved reflection change as a function of $\omega$: (d) $\Delta R_{eo}(\omega,\delta)/R_0$ obtained by the simulation
  and (e) $\Delta R_{eo}(\omega,\delta)/R_0(\omega)$ by simulations and analytical calculations (Eqs.(\ref{dReo(w)R0(w)}) and (\ref{dR-w-a})). 
}
\label{fig-dR}
\end{figure}

We next consider the modulation in the reflectivity.
In the first-principles calculation, the reflected wave in the vacuum region left to the surface
is composed of that by the direct reflection at the front surface ($X=$0 $\mu$m) and that by the reflection
at the back surface ($X=$6 $\mu$m) after the propagation inside the medium. The latter component includes
stimulated Raman wave while the former does not.
In the first-principles calculation, the former reflected wave contains extremely weak $\hat z$-component,
while the latter reflected wave accompanying substantial $\hat z$-component
that comes from the stimulated Raman wave.
We first discuss the contribution of the former process without the propagation inside the medium.

The reflection change is shown in Fig.\ref{fig-dR}(a) by dots for three different pump-probe delay time
of $\delta$ = 83.0, 89.5, and 96.0 fs.
The modulation in the reflection calculated by the first-principles calculation coincides accurately 
with the analytics formula of Eq. (\ref{dR}) that shows a sine-like dependence.
In panels (b) and (c), the frequency-dependent modulation at $\omega_0 \pm \Omega$ is shown.
They shows a similar sine-like behavior and are again well reproduced by the analytic formula.
Therefore, the validity of the analytic formula is confirmed with high accuracy for the reflected wave.

In panels (d) and (e), we show modulation of the reflectivity in frequency domain.
The frequency-resolved modulation divided by the reflectivity with/without frequency resolution
is shown in the panel (e)/(d), respectively.
As seen in (d), the modulation has a similar frequency dependence with the frequency-resolved flux
of the incident wave.
The frequency-resolved reflectivity shown in (e) indicates that the frequency-dependence of the
modulation is rather weak. 
In the frequency region far apart from the central frequency, the modulation becomes larger.
However, the incident flux does not have much component in such frequency region. 
We find a good agreement among three curves, the first-principles calculation, the analytic formula
of Eq. (\ref{dReo(w)R0(w)}), and the simplified analytic formula of Eq. (\ref{dR-w-a}). 
The agreement indicates that the analytic formula are sufficiently accurate to describe the modulation
in the reflectivity.

\subsection{Reflection at the back surface}
\vspace{-3mm}

\begin{figure}[ht]
\centering
\includegraphics[angle=0,width=0.98\linewidth]{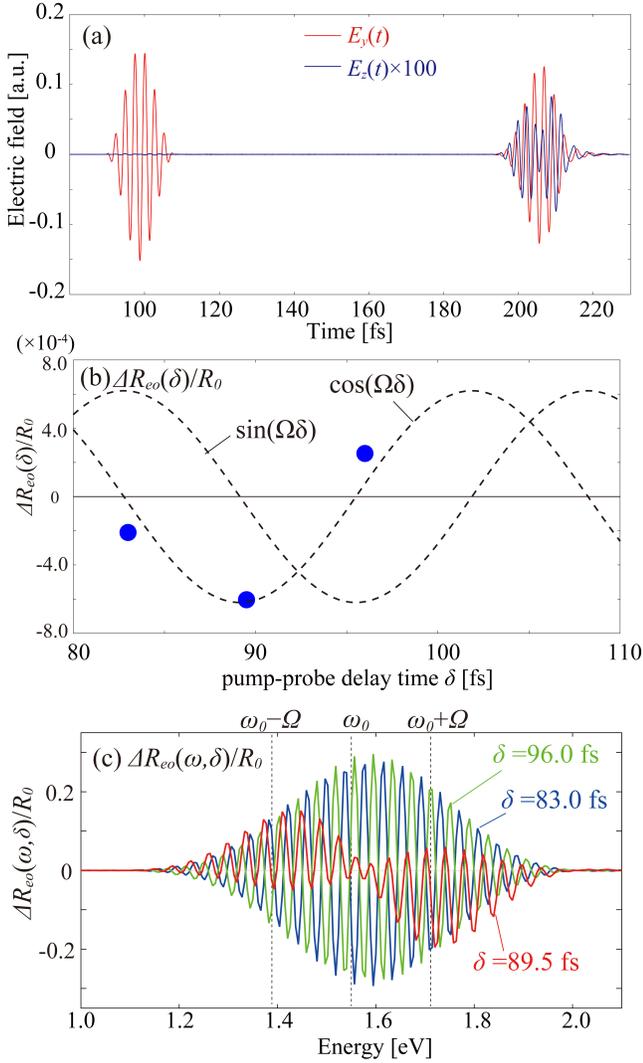}
\vspace{-4mm}
\caption{
  (a) Example of the reflection wave from the front surface and the back side detected at the left vacuum region ($\delta$=89.5 fs).
  (b) Reflection change as a function of $\delta$ taking into account of the reflection waves from both surfaces, and
  (c) their spectrally resolved reflection changes. 
}
\label{fig-dR-back}
\end{figure}

We next examine the modulation in the reflection including the reflected wave
caused by the back-surface. 
In Fig.\ref{fig-dR-back}(a), we show a time profile of the reflection waves reflected 
from the front surface and from the back surface of the medium in the first-principles calculation.
The $z$-component of the field is very small in the first wave from the front surface, 
while that of the second wave from the back side is much larger because of the amplification 
of the Raman wave during the propagation.
The reflection change $\Delta R_{eo}(\delta)/R_0$ is shown in Fig. \ref{fig-dR-back}(b) for three
pump-probe delay times, $\delta$=83.0, 89.5, and 96.0 fs.
It can be well fit by a cosine-like function.
This can be understood as follows:
In the case of the transmission, the bulk effect caused by the stimulated Raman wave is
dominated as seen in Fig.\ref{fig-dT}(a). In the present case, the second reflected wave includes the similar Raman wave
component as seen in the panel (a). The modulation in the reflection is dominated by the
bulk effect, if we include the second reflected wave at the back surface.

Fig. \ref{fig-dR-back}(c) shows the spectrally resolved signals calculated by using the first and second reflection waves.
A strong oscillation structure is observed in the frequency domain for three cases of the time delay.
The oscillation structure is due to the interference between the first and the second waves:
The Fourier transformed electric field of the reflection wave can be given as
$\widetilde{E}_1(\omega)+\widetilde{E}_2(\omega)e^{i\omega (t_2-t_1)}$,
where $\widetilde{E}_1$ and $\widetilde{E}_2$ are of the first and the second reflection waves, respectively,
and $t_1$ and $t_2$ is the arrival time of the first and second waves, respectively.
The power spectrum is written as
$I(\omega)=|\widetilde{E}_1(\omega)|^2+|\widetilde{E}_2(\omega)|^2+2{\rm Re}\left[ \widetilde{E}_1^*(\omega)\widetilde{E}_2(\omega)e^{i\omega (t_2-t_1)}\right]$.
It indicates that the oscillation frequency is inversely proportional to the difference of the two
reflected waves, $t_2-t_1$, which is proportional to the thickness of the sample.
The present calculation assumes a sample of 6 $\mu$m thickness.
The oscillation will not be observed if a much thicker sample is utilized.

If we average the signals of Fig. \ref{fig-dR-back}(c) over rapidly oscillating structure,
we obtain a very small signal for $\delta$=83.0 and 96.0 fs, and a strong signal remains
for $\delta$=89.5 fs. The averaged feature is very close to the transmission shown in Fig.\ref{fig-dT}(d).
Namely, the signal is caused mainly by the Raman wave that is included in the second
wave of Fig.\ref{fig-dR-back}(a) and that is quite similar to that in the transmitted wave.

\section{Summary}   \label{sec:summary}

We have presented a comprehensive theoretical analysis on the probe stage of pump-probe measurements
of coherent phonon generation in dielectrics.
We take a diamond as a typical case and assume the impulsive stimulated Raman scattering mechanism
for the generation process.

We have developed analytical and computational approaches.
In analytical description,
we revisited the work developed in Ref. [\onlinecite{Merlin1997}] by Merlin and developed comprehensive formula. 
We start with a standard description of light propagation coupled with 
a phonon motion through the Raman tensor.
We summarize formula for the modulation on the reflection and transmission of the probe pulse
using a perturbative solution for the probe pulse.
The modulation in the transmission is caused by two distinct mechanisms: the boundary and the
bulk effects. The bulk effect is caused by the stimulated Raman wave that is amplified as the
probe pulse propagates in the medium.
The modulation in the reflection is caused by the boundary effect. However, if we consider the
reflection at the back surface, the bulk effect also contribute in the reflection.

The modulation is investigated for frequency-resolved and -integrated signals.
The boundary and the bulk effects contribute to the modulation in qualitatively different way.
The bulk effect produces strong frequency dependence in the modulation, whereas the boundary
effect produces very weak frequency dependence.
The bulk effect causes strong modulation in the probe signal when the probe pulse moves
with the nodal point of the phonon. It causes a phase shift of $\pi/2$ between the phonon
amplitude and the probe signal.
Contrarily, the boundary effect causes a modulation that is proportional to the amplitude of
the phonon. The modulation of the probe pulse is in phase with the coherent phonon.

The derivation of the analytic formula is based on several assumptions and approximations.
To confirm the validity of the analytic formula, we performed first-principles calculations
based on time-dependent density functional theory. In our multiscale formalism, coupled dynamics
of mesoscopic light propagation and microscopic electronic and ionic motions are described
simultaneously without any empirical parameters.

By comparing results between the analytical theory and the first-principles calculation,
we confirmed the validity and the reliability of the analytical formula.
We thus consider that our analytic formula provides a reliable basis for the experimental
analysis of frequency-resolved modulation in the pump-probe measurement of coherent phonon
in transparent dielectrics.

\section{Acknowledgement}
We thank Professor K. G. Nakamura for useful discussion. 
We acknowledge the supports by JST-CREST under grant number JP-MJCR16N5, and 
by MEXT as a priority issue theme 7 to be tackled by using Post-K Computer, 
and by JSPS KAKENHI Grant Number 15H03674. 
Calculations are carried out at Oakforest-PACS at JCAHPC through the Multidisciplinary 
Cooperative Research Program in CCS, University of Tsukuba, and through the HPCI System 
Research Project (Project ID: hp180088).

\bibliographystyle{unsrt}
\bibliography{bibf/yabana,bibf/theory,bibf/laser,bibf/tddft,bibf/coherent_phonon,bibf/other_misc_1}

\end{document}